\documentclass[twocolumn,separability,Amsterdam, nobibnotes, aps, pra,showpacs,preprintnumbers]{revtex4-1}
\usepackage{amssymb}
\usepackage{array}
\usepackage{graphicx}

\def\beq{\begin{equation}}
\def\eeq{\end{equation}}
\def\bea{\begin{eqnarray}}
\def\eea{\end{eqnarray}}

\usepackage{color}

\begin{document}

\title{  Entanglement Thermodynamics of the Generalized Charged BTZ Black Hole}%

\author{Seyed Ali Hosseini Mansoori,$^{1,2,3}$ Behrouz Mirza,$^2$ Mahdi Davoudi Darareh,$^3$ and Sharooz Janbaz$^3$ }
\affiliation{$^1$ Department of Physics, Boston University, 590 Commonwealth Ave., Boston, MA 02215, USA\\
$^2$Department of Physics, Isfahan University of Technology, Isfahan 84156-83111, Iran\\
$^3$Faculty of Applied Sciences, Malek Ashtar University of Technology, P.O.Box 115/83145, Isfahan, Iran}
\email{shossein@bu.edu; \,\,\ \\
b.mirza@cc.iut.ac.ir; \,\,\ \\
m.davoudi@mut-es.ac.ir; \,\,\ \\
shjanbaz@mut-es.ac.ir}
\date{\today}%

\begin{abstract}
In this paper, we investigate the entanglement entropy for the generalized charged BTZ black hole through the $AdS_{3}/CFT_{2}$ correspondence. Using the holographic description of the entanglement entropy for the strip-subsystem in boundary $CFT_{2}$, we will find the first law-like relation between the variation of holographic entanglement entropy and the variation of energy of the subsystem in terms of the mass and the electric charge up to the second order. We also obtain appropriate counterterms to renormalize the energy tensor associated with the bulk on-shell actions.
\end{abstract}


\maketitle

\section{Introduction}\label{a}
The anti-de Sitter space and conformal field theory (AdS/CFT) correspondence states a duality between quantum field theory and gravity \cite{mal,mal2}.
More precisely, the correspondence establishes a relationship between  gravitational theories on $AdS$ spacetime and a quantum field theory that lives on the conformal boundary of the $AdS$ spacetime. This duality is an example of a more general theory named  "the holographic principle" \cite{gt, witten, gubser}. The holographic principle emerged from the study of the black hole entropy \cite{lbo, lbo2}.

 In recent years, AdS/CFT has been
the subject of much research in areas other than merely
 theoretical particle physics or string theory. It has been shown to serve as  a powerful tool for analyzing a natural phenomena in a variety of fields, ranging from QCD, nuclear physics to nonequilibrium physics, and condensed-matter physics \cite{jca, saha, niq, niq2}.
In spite of its many applications in different areas of physics, the fundamental mechanism
of the AdS/CFT correspondence is still far from clear.

The AdS/CFT duality has also been  used to make a geometric model for evaluating the entanglement of quantum systems
using conformal field theory descriptions. One of the main developments in this direction is
the suggestion by Ryu and Takayanagi \cite{sry, sry2,sry3, sry4} according to which  the entanglement entropy (EE) of a CFT's  states living on the boundary of an AdS spacetime is associated with the area of a minimal surface defined in the bulk of that spacetime. Intuitively, the minimal surface plays the key role in the holographic
principle for an observer who is only accessible to an explicit subsystem of a system \cite{sry, sry2,sry3, sry4}. Originally,  Ryu and Takayanagi's proposal was deduced  from the Bekenstein-Hawking entropy \cite{je, hca, dvf,mca, rem}.

Over the last few years, the entanglement thermodynamics, which includes  EE as the Bekenstein-Hawking entropy, has attracted a lot of attention \cite{jbh, she, dal, fca, lma, wzg, afa, aem}. For instance, the authors in \cite{jbh} investigated the low thermal excited state from a holographic viewpoint.  The relation between the variance of  energy, $\Delta E$, and the variance of EE, $\Delta S$, for the subsystem in the low excited states of CFT is similar to the first law of thermodynamics; i.e., $\Delta E={{T}_{ent}}\Delta S$, where the entanglement temperature, $T_{ent}$, has been identified with the inverse of the size of the entangling region. Moreover, the Holographic entanglement entropy (HEE) of the boundary $CFT$ with low-energy excited states up to the second order can be expressed by the first law-like thermodynamic relation when the spatial region of the boundary subsystem is a strip \cite{she}.

In this paper, we will study the properties of the quantum entanglement of the generalized charged BTZ black hole using the
AdS/CFT duality. The generalized charged BTZ solutions are derived by considering three types of  nonlinear
electromagnetic fields (NLED) coupled with Einstein gravity  \cite{shendi}.
Becuase these solutions have asymptotical behaviors similar to the AdS solution at radii much larger than the radius of the horizon, these solutions can be considered as the perturbed geometry, deviating from the pure AdS spacetime via small perturbations introduced by the mass and the electic charge. Following a methodology similar to \cite{jbh, she}, we derive the
relation between the holographic entanglement entropy and the energy at the boundary up to second order in terms of the mass and the electric charge of the generalized charged BTZ black hole.
Since the bulk Einstein equations and the equation of motion for gauge fields hold for all orders of the small perturbation, we expect that the first law-like relation to be also satisfied for the second order in which the entanglement temperature gets modified. In addition, the second order energy of the subsystem can be interpreted as the terms responsible for nonzero expectation value of the stress tensor in the CFT side, which is related to the on-shell gravity action. However, this term diverges. To regulate the theory we get rid of all infinities by adding counterterms and finally remove the regulator to obtain the renormalized action \cite{vbala, kostas}.
\newpage
The paper is organized as follows. In the next Section, we will have a brief glance at the generalized charged BTZ solutions. In Sections \ref{S3} and \ref{S4}, we will calculate the second order HEE and the second order stress tensor of the boundary before proceeding to analyze the first law-like
thermodynamics at second order. Finally, Section \ref{S5} is devoted to conclusions.
\section{generalized charged BTZ black hole}\label{S1}
 In this section, we begin with a breif review of black hole solutions in Einstein
gravity with a negative cosmological constant coupled with NLED fields. 
The action of Einstein gravity with NLED field \cite{shendi} is given
by:
\begin{equation}\label{1}
S_{grav}=S_{EH}+S_{gauge}+{{S}_{GH}}
\end{equation}
where, the Einstein-Hilbert action, $S_{EH}$, the gauge field action, $S_{gauge}$, and the Gibbons-Hawking action, $S_{GH}$ respectively, are given by the following expressions:
\begin{eqnarray}\label{2}
 S_{EH}&=&-\frac{1}{16\pi }\int_{M}{{{d}^{3}}x\sqrt{-g}}\left[ R+\frac{2}{L^2} \right]\\
 S_{gauge}&=&-\frac{1}{16\pi }\int_{M}{{{d}^{3}}x\sqrt{-g}}L(F)\label{w1}\\
 {{S}_{GH}}&=&-\frac{1}{8\pi }\int_{\partial M}{{{d}^{2}}x\sqrt{-h }}K
\end{eqnarray}
where, $L$, $R$, and $L(F)$ are the AdS length, the Ricci scalar for the bulk manifold, $M$, and the Lagrangian of NLED field,
respectively. Here, $h$ is the determinant of the 2-dimensional metric at the boundary of manifold $M$ ($\partial M$), and $K$ is the trace of the extrinsic curvature of the boundary. We will examine the case where the Lagrangian of the NLED field, $L(F)$ is the Born-Infeld nonlinear electromagnetic (BINEF), the exponential form of nonlinear electromagnetic field (ENEF), and the logarithmic form of nonlinear electromagnetic field (LNEF) \cite{shendi}:
\begin{equation}\label{3}
L(F)=\left\{ \begin{array}{ll}
\noalign{\medskip}   4{{\beta }^{2}}\left( 1-\sqrt{1+\frac{F}{2{{\beta }^{2}}}} \right) & BINEF \\
\noalign{\medskip}   {{\beta }^{2}}\left( \exp \left( -\frac{F}{{{\beta }^{2}}} \right)-1 \right) & ENEF \\
\noalign{\medskip}   -8{{\beta }^{2}}\ln \left( 1+\frac{F}{8{{\beta }^{2}}} \right) & LNEF \\
\end{array} \right.
\end{equation}
where, $\beta$ and $F = F_{\mu \nu}F^{\mu \nu}$ are the nonlinearity parameter and the Maxwell invariant, respectively. By varying Eq. (\ref{1}) with respect to the gravitational field, $g_{\mu \nu}$ and the gauge field, $A_{\mu}$, we have:
\begin{eqnarray}\label{4}
&{{R}_{\mu \nu }}-\frac{1}{2}{{g}_{\mu \nu }}\left( R+\frac{2}{L^{2}}  \right)=\frac{1}{2}{{g}_{\mu \nu }}L(F)-2{{F}_{\mu \sigma }}F_{\nu }^{\sigma }\frac{dL(F)}{dF}\\
&{{\partial }_{\mu }}\left( \sqrt{-g}\frac{dL(F)}{dF}{{F}^{\mu \nu }} \right)=0\label{e5}
 \end{eqnarray}
By substituting the gauge potential, ${{A}_{\mu }}=\varphi (r)\delta _{\mu }^{0}$, in the NLED fields in Eq. (\ref{e5}), it is easy to obtain the non-vanishing components of the electromagnetic field tensor as follows:
\begin{equation}
{{F}_{tr}}=\frac{q}{r}\times \left\{ \begin{array}{cc}
   {{\Gamma }^{-1}} & BINEF  \\
   \frac{r\beta \sqrt{{{L}_{w}}}}{2q} & ENEF  \\
   \frac{2{{\beta }^{2}}{{r}^{2}}}{{{q}^{2}}}\left( \Gamma -1 \right) & LNEF   \end{array} \right.
\end{equation}
where $\Gamma =\sqrt{1+\frac{{{q}^{2}}}{{{r}^{2}}{{\beta }^{2}}}}$ and ${{L}_{w}}=LambertW\left( \frac{4{{q}^{2}}}{{{r}^{2}}{{\beta }^{2}}} \right)$ that satisfies $LambertW(x)\exp[LambertW(x)] = x$. Moreover, Eq. (\ref{4}) admits a solution of the following form.

\begin{equation}\label{6}
d{{s}^{2}}=-\frac{{{r}^{2}}g(r)}{{{L}^{2}}}d{{t}^{2}}+\frac{{{L}^{2}}d{{r}^{2}}}{{{r}^{2}}g(r)}+{{r}^{2}}d{{\theta }^{2}}
\end{equation}
where,
\begin{widetext}
\begin{equation}\label{7}
g(r)=1-\frac{{{L}^{2}}M}{{{r}^{2}}}+\left\{ \begin{array}{ll}
   2{{L}^{2}}{{\beta }^{2}}(1-\Gamma )+\frac{{{q}^{2}}{{L}^{2}}}{{{r}^{2}}}\left[ 1-2\ln \left( r\frac{(1+\Gamma )}{2L} \right) \right] & BINEF  \\
   \frac{\beta q{{L}^{2}}\left( 1-2{{L}_{w}} \right)}{r\sqrt{{{L}_{w}}}}-\frac{{{\beta }^{2}}{{L}^{2}}}{2}+{{\frac{{{q}^{2}}L}{{{r}^{2}}}}^{2}}\left[ \ln \left( \frac{{{\beta }^{2}}{{L}^{2}}}{2{{q}^{2}}} \right)-Ei\left( 1,\frac{{{L}_{w}}}{2} \right)-\gamma +3 \right] & ENEF  \\
   4{{\beta }^{2}}{{L}^{2}}\left[ \ln \left( \frac{\Gamma +1}{2} \right)+3 \right]-{{\frac{{{q}^{2}}L}{{{r}^{2}}}}^{2}}\left[ \ln \left( \frac{{{\beta }^{2}}{{r}^{4}}\left( \Gamma -1 \right){{\left( \Gamma +1 \right)}^{3}}}{4{{q}^{2}}{{L}^{2}}} \right)+\frac{6}{\Gamma -1}-2 \right] & LNEF  \\ \end{array} \right.
\end{equation}
\end{widetext}
in which $M$ and $q$, respectively, are the mass and the electric charge of the generlized charged BTZ black hole. Furthermore $\gamma =\gamma (0)=0.57722$, and the special function $Ei(1,x)=\int_{1}^{\infty }{\frac{{{e}^{-xz}}}{z}dz}$.

 \section{ the Holographic Entanglement Entropy } \label{S3}
Now, we study the holographic entanglement entropy of $CFT_2$ with second order excitations in the generalized charged BTZ black hole by writing the metric as pure $AdS_3$ plus small metric perturbations in the asymptotic limit. Therefore we consider a quantum field theory defined on a 2-dimensional manifold consisting of the time direction and a 1D spacelike manifold. At a fixed time $t=t_0$, the spacelike manifold is then broken down into two subsystems,  $A$ and its complement $B$. The entanglement entropy of subsystem $A$ is defined as the von Neumann entropy ${{S}_{ent}}(A)=-\mathrm{Tr}\left[ {{\rho }_{A}}\ln {{\rho }_{A}} \right]$ related to the reduced density matrix ${{\rho }_{A }}=\mathrm{Tr}_{B}{{\rho }_{tot}}$, obtained by taking a trace of the density matrix $\rho_{tot}$ for the total system over the subsystem $B$ \cite{pcal, pjca, jcal, sns}. But in the gravity dual, Ryu and Takayanagi have calculated the holographic entanglement entropy \cite{sry, sry2, sry3} to be:
\begin{equation}\label{q2}
{{S}_{\gamma_{A}}}=\frac{A_{{\gamma }_{A}}}{4{{G}_{3}}}
\end{equation}
where,  $\gamma_{A}$ is the one dimensional static minimal surface in $AdS_{3}$ whose boundary coincides with $\partial A$. Now, we carry out the actual calculation of entanglement entropy for the generalized charged BTZ black hole. By defining $u={{L}^{2}}/r$, and $x=L\theta $ in the Poincare's coordinate, and replacing them into Eq. (\ref{6}), we have:
\begin{equation}\label{q3}
d{{s}^{2}}=\frac{{{L}^{2}}}{{{u}^{2}}}\left[ -g(u)d{{t}^{2}}+\frac{d{{u}^{2}}}{g(u)}+d{{x}^{2}} \right]
\end{equation}
where $g(u)$ function obtained as Eq. (\ref{77}). In this metric function, we have defined $\kappa =\frac{M}{{{L}^{2}}}$ and $\lambda =\frac{q}{L}$. Now, let us consider an entangling region (subsystem A) in the shape of a strip with the width of $x\in \left[ -\frac{\gamma }{2},\frac{\gamma }{2} \right]$ \cite{sry2}.
\begin{widetext}
\begin{equation}\label{77}
g(u)=1-\kappa {{u}^{2}}+\left\{ \begin{array}{ll} 
   2{{L}^{2}}{{\beta }^{2}}(1-\Gamma )+{{\lambda }^{2}}{{u}^{2}}\left[ 1+2\ln \left( \frac{2u}{L(1+\Gamma )} \right) \right] & BINEF  \\
   \frac{\beta qu\left( 1-2{{L}_{w}} \right)}{\sqrt{{{L}_{w}}}}-\frac{{{\beta }^{2}}{{L}^{2}}}{2}+{{\lambda }^{2}}{{u}^{2}}\left[ \ln \left( \frac{{{\beta }^{2}}}{2{{\lambda }^{2}}} \right)-Ei\left( 1,\frac{{{L}_{w}}}{2} \right)-\gamma +3 \right] & ENEF  \\
   4{{\beta }^{2}}{{L}^{2}}\left[ \ln \left( \frac{\Gamma +1}{2} \right)+3 \right]+{{\lambda }^{2}}{{u}^{2}}\left[ \ln \left( \frac{4{{\lambda }^{2}}{{u}^{4}}}{{{\beta }^{2}}{{L}^{4}}\left( \Gamma -1 \right){{\left( \Gamma +1 \right)}^{3}}} \right)-\frac{6}{\Gamma -1}+2 \right] & LNEF  \\ \end{array} \right.
\end{equation}
\end{widetext}
The minimal surface $\gamma_{A}$ is a 1-dimensional hypersurface (geodesic) at $t = 0$ when  Eq. (\ref{q2}) is employed. It should be noted  that none of the coordinates $(u; x)$ is independent of the other. Therefore, considering $u$ as a function of $x$, the surface area becomes:
\begin{equation}\label{q1}
A_{{\gamma }_{A}}=2L\int_{-\frac{\gamma}{2}}^{\frac{\gamma}{2}}{\frac{dx}{u(x)}\sqrt{\frac{\left( {{\partial }_{x}}u(x) \right)^{2}}{g(u)}+1}}
\end{equation}
It is noteworthy that we have not imposed the minimality condition on the surface area yet. The minimal
surface should be stable against a small perturbation, $u(x) \to u(x)+\delta u(x)$, and should thus satisfy the equation of motion.
\begin{equation}
\frac{\delta {{A}_{{{\gamma }_{A}}}}}{\delta u(x)}=0
\end{equation}
The above equation gives us a second order differential equation. Although it may be solved by imposing the Dirichlet boundary conditions $\delta u(x) \to 0$ and the boundary condition $u\to \varepsilon $ at $x=\pm \frac{\gamma}{2}$ ( $\epsilon$ is the UV cutoff), it is
easier to think of the integrand of (\ref{q1}) as a Lagrangian with $x$ being the time parameter. From the Hamiltonian we get:
\begin{equation}\label{eqr1}
H=-{{\left[ u\sqrt{\frac{{{\left( {{\partial }_{x}}u \right)}^{2}}}{g(u)}+1} \right]}^{-1}} \,\ ; \,\ \frac{\partial H}{\partial x}=0
\end{equation}
Denoting ${{u}^{*}}$ as the maximum value of $u(x)$, where ${{\partial }_{x}}u(x)=0$, one could determine the constant, $H=-1/{{u}^{*}}$ \cite{sry2}. Thus form relation (\ref{eqr1}), the first order differential equation can be written as,
\begin{equation}\label{q7}
\partial_{x}u=\sqrt{g(u) \left(\left( \frac{{u}^{*}}{u} \right)^{2}-1 \right)}.
\end{equation}
By integrating $dx$ from $0$ to $\frac{\gamma}{2}$, we have:
\begin{equation}\label{q5}
\frac{\gamma }{2}=\frac{{{u}^{*}}}{2}\int_{0}^{1}{\frac{d\xi }{\sqrt{g(\xi )\left( 1-\xi  \right)}}}
\end{equation}
where, $\xi ={{\left( \frac{u}{{{u}^{*}}} \right)}^{2}}$. This expression represents the relation between the width $\frac{\gamma}{2}$ and the maximum $u^{*}$. On the other hand, by substituting Eq. (\ref{q7}) into Eq. (\ref{q1}), the minimal surface area becomes:
\begin{equation}\label{au3}
A_{\gamma_{A}}=L\int_{\left( \frac{\varepsilon }{{{u}^{*}}} \right)^{2}}^{1}{\frac{d\xi }{\xi \sqrt{g(\xi )(1-\xi )}}}
\end{equation}
Now, let us consider Eq. (\ref{au3}) as a slightly perturbed geometry obtained from the
pure AdS spacetime. Therefore, in order to determine the shape of the bulk minimal surface, one
needs to consider $\kappa$ and $\lambda$ as small parameters acting as the sources of the geometric perturbation away from the pure $AdS_{3}$ spacetime \cite{she}. By rewriting Eq. (\ref{77}) in the new coordinate $\xi$ and then expanding the function $1/ \sqrt{g(\xi)}$ in Eq. (\ref{q5}) up to the second order with respect to $\kappa$ and $\lambda $,  the $u^{*}$ can be calculated up to the second order as follows:
\begin{eqnarray}\label{eu}
 {{u}^{*}}&\approx &{{u}^{*}}^{\left( 0 \right)}+{{u}^{*}}^{\left( 1 \right)}+{{u}^{*}}^{\left( 2 \right)}=\frac{\gamma }{2}-\frac{1}{3}{{\left( \frac{\gamma }{2} \right)}^{2}}\kappa \\
\nonumber &+&\frac{2}{15}{{\left( \frac{\gamma }{2} \right)}^{5}}{{\kappa }^{2}}-\frac{x_{1}{{\lambda }^{2}}{{\gamma }^{3}}}{12}\left[ \ln \left( \frac{L}{\gamma } \right)-\frac{y_{1}}{6} \right]\text{ }
\end{eqnarray}
where $x_{1}=\left\{ 0.5,1,1 \right\}$ and $y_{1}=\left\{7, 1, 5\right\}$. (In the following, the symbol $\left\{...\right\}$ is used for variable values for BINEF, ENEF, and LNEF fields, respectively.)
According to  Eqs. (\ref{q2}), (\ref{au3}), and Eq. (\ref{eu}), the HEE of the subsystem A is given by:
\begin{eqnarray}\label{s1}
&\nonumber {{S}_{{{\gamma }_{A}}}} \simeq  S_{{{\gamma }_{A}}}^{\left( 0 \right)}+S_{{{\gamma }_{A}}}^{\left( 1 \right)}+S_{{{\gamma }_{A}}}^{\left( 2 \right)}+ O\left( {{\kappa }^{3}},{{\lambda }^{3}},\kappa {{\lambda }^{2}},{{\varepsilon }^{2}} \right)=\frac{c}{3}\ln \left( \frac{\gamma }{\varepsilon } \right)\\
&+\frac{c\kappa }{18}{{\left( \frac{\gamma }{2} \right)}^{2}}-\frac{c{{\kappa }^{2}}}{540}{{\left( \frac{\gamma }{2} \right)}^{4}}-\frac{x_{2}c{{\lambda }^{2}}}{18}{{\left( \frac{\gamma }{2} \right)}^{2}}\left[ \frac{y_{2}}{3}+\ln \left( \frac{\gamma }{L} \right) \right]
\end{eqnarray}
where $x_{2}=\left\{ 2,1,1 \right\}$ and $y_{2}=\left\{2, 1, 4\right\}$ and $c=3L/2G_{3}$ is the central charge of the boundary \cite{jdbro}. Note that the first term in the above expantion is the vacuum entanglement entropy which shows the UV divergence at boundary ($\epsilon \to 0$). In the next section, we only consider the renormalized entanglement entropy by ignoring the vacuum state from the HEE expansion.
\section{First law of the entanglement thermodynamics}\label{S4}
In this section, we obtain the excitation of energy levels by extracting the Brown-York tensor at the CFT boundary. Moreover, using the pervious results for HEE and new expressions for energy, we try to write the first law-like thermodynamics for each order by modifying the entanglment temperature .

 The metric Eq. (\ref{q3}) can be recast in the ADM form as follows:
 \begin{equation}
 d{{s}^{2}}={{g}_{uu}}du^{2}+{{h}_{\mu \nu }}d{{x}^{\mu }}d{{x}^{\nu }}
 \end{equation}
 in which, ${{h}_{\mu \nu }}$ is the induced metric on the $u$ hypersurface. Then, the extrinsic curvature of a fixed $u$ hypersurface reads:
 \begin{equation}
{{K}_{\mu \nu }}=\frac{1}{2}{{n}^{u}}{{\partial }_{u}}{{h}_{\mu \nu }}
 \end{equation}
where ${{n}^{u }}=\frac{1}{\sqrt{{{g}_{uu}}}}$ is the unit normal vector on the $u$ constant hypersurface. Moreover, the extrinsic curvature is defined by $K =h^{\mu \nu}{{K}_{\mu \nu }}$.  The renormalized stress tensor is obtained as follows:
\begin{equation}\label{au6}
T_{\mu \nu }^{ren}=\frac{1}{8\pi G}\left[ {{K}_{\mu \nu }}-K{{h}_{\mu \nu }} \right]+T_{\mu \nu }^{ct}
\end{equation}
 where $T_{\mu \nu }^{ct}$ is the counterterm tensor which is added to the Brown-York tensor in order to obtain a finite stress
tensor \cite{jdborn, vbala}.  For our problem, the counterterm tensor consists of two actions. The first action comes from a local functional of the intrinsic geometry of the boundary chosen to remove the divergences that arise as $u$ hypersurface tends to the $AdS_{3}$ boundary. Because the generalized charge BTZ solutions behave like the $AdS_{3}$ space at the boundary ($u \to 0$), the counterterm is just the boundary cosmological constant \cite{vbala}.
\begin{equation}\label{q14}
{{S}_{ct}^{1}}=-\frac{1}{8\pi GL}\int_{\partial {{M}}}{d{{x}^{2}}\sqrt{-h}}
\end{equation}
And the other counterterm action is $S_{ct}^{F}$ which cancels the divergences from the gauge field action at the boundary  \cite{vbala, kjensen, kostas}. In order to determine $S_{ct}^{F}$, we consider the expantion of the $L(F)$'s defined in Eq. (\ref{3}) for large values of $\beta$. Therefore, the guage field action Eq. (\ref{w1}) can be rewritten as:
\begin{equation}
{{S}_{gauge}}={{\sum\limits_{n}{{{c}_{n}}{{\beta }^{2\left( 1-n \right)}}\int_{M}{{{d}^{3}}x\sqrt{-g}}\left[ {{F}_{\mu \nu }}{{F}^{\mu \nu }} \right]}}^{n}}
\end{equation}
where $c_{n}$ are expansion coefficients \cite{shendi}. By expressing the bulk field strength $F_{\mu \nu}$ in the dual form
of the scalar field $\varphi$  by ${{F}^{\mu \nu }}={{\varepsilon }^{\mu \nu \sigma }}{{\partial }_{\sigma }}\varphi /\sqrt{-g}$, we have:
\begin{equation}
{{S}_{gauge}}={{\sum\limits_{n}{2{{{c}_{n}{{\beta }^{2\left( 1-n \right)}}}}}\int_{M}{{{d}^{3}}x\sqrt{-g}}\left[ \partial \varphi  \right]}}^{2n}
\end{equation}
So the bulk field equation for each $n$ is thus:
\begin{eqnarray}
 {{\nabla }_{\mu }} \left( {{\partial }^{\mu }}\varphi {{\left[ \partial \varphi  \right]}^{2n-2}} \right)=0
\end{eqnarray}
To regularize this action, we divide the range of the $u$ integration into two areas: the bulk term, $u\ge \epsilon $, and the
boundary terms at $u=\epsilon$ where $\epsilon$ is the cutoff \cite{kostas}.
 \begin{eqnarray}\label{q12}
& \nonumber  {{S}_{reg}}\simeq \sum\limits_{n}{2{{c}_{n}}{{\beta }^{2\left( 1-n \right)}}}[\int_{u\ge \varepsilon }{{{d}^{3}}x \sqrt{-g}\varphi}{{\nabla }_{\mu }} \left( {{\partial }^{\mu }}\varphi {{\left[ \partial \varphi  \right]}^{2n-2}} \right)\\
& -\int_{u=\varepsilon }{{{d}^{2}}x\sqrt{-g}}\varphi {{\partial }^{\mu }}\varphi {{\left[ \partial \varphi  \right]}^{2n-2}}]=\\
&\nonumber -\sum\limits_{n}{2{{c_{n}}{{\beta }^{2\left( 1-n \right)}}}}[{\int_{u=\varepsilon }{d{{x}^{2}}\sqrt{{{g}_{uu}}}\sqrt{-h}}{{g}^{uu}}{{\left( \partial \varphi  \right)}^{2n-2}}\varphi {{\partial }_{u}}\varphi }]
    \end{eqnarray}
 In the last term, since the bulk field equations are satisfied, the bulk term vanishes. Near the boundary, the diagonal metric $h_{\mu \nu}$ has the same asymptotic expansion as the Fefferman-Graham series \cite{feff},
\begin{eqnarray}\label{qq11}
&{{h}_{\mu \nu }}\simeq {{u}^{-2}}h_{\mu \nu }^{\left( 0 \right)}+h_{\mu \nu }^{\left( 2 \right)}+{{u}^{2}}h_{\mu \nu }^{\left( 4 \right)}+...\\
&\nonumber \sqrt{-h}\simeq \sqrt{{{-h }^{(0)}}}\left[ {{u}^{-1}}+\frac{\mathrm{Tr}{{h }^{(2)}}}{2}+... \right]
\end{eqnarray}
where, $\mathrm{Tr}{{h }^{(2)}}={{h }^{(0)\mu \nu }}h _{\mu \nu }^{(2)}$. These are asymptotically $AdS_{3}$ geometries with a boundary at $u \to 0$.
Furthermore, the explicit asymptotic solution for the scalar field $\varphi$ near the boundary takes the following form \cite{kjensen, kostas}
\begin{equation}\label{q11}
\varphi (u ,x)={{\varphi }^{(0)}}(x)+u^{2} \left[ {{\varphi }^{(2)}}(x)-\frac{{L^{2}}}{2}{{\square }_{{{h }^{\left( 0 \right)}}}}{{\varphi }^{(0)}}(x)\ln \left( \frac{u}{L} \right) \right]
\end{equation}
 where, ${{\square }_{h^{0}}}$ stands for the Laplacian operator made by the induced metric $h _{\mu \nu }^{\left( 0 \right)}=u {{h}_{\mu \nu }}$ at $u \to \epsilon$ and the functions $\varphi^{(0)}$ and $\varphi^{(2)}$ indicate boundary
data for the scalar field \cite{kjensen}. Substituting Eq. (\ref{q11}) and Eq. (\ref{qq11}) in Eq. (\ref{q12}), we can show that the gauge action is logarithmically divergent only for $n=1$ when $\varphi^{0}$ has a gradient, while for $n\ne 1$, there is no divergent expression. We will, therefore, have:
\begin{equation}
S_{ct}^{\varphi } \propto \int_{u = \epsilon} {d^{2}{{x}}\sqrt{-h}}{{\left( \partial \varphi  \right)}^{2}}\ln \left(\frac{u}{L} \right)
\end{equation}
Moreover,  we are able to state this counterterm in terms of $F_{\mu \nu}$ in the original coordinates as:
\begin{equation}\label{q13}
{S_{ct}}^{F}=-{{\alpha}}\int_{M} {d{{x}^{2}}\sqrt{-h}{{F}_{r\mu }}{{F}^{r\mu }}\ln \left( \frac{r}{L} \right)}
\end{equation}
Therefore, the stress tensor associated with the two counterterms, Eq. (\ref{q14}) and Eq. (\ref{q13}) is given by:
\begin{eqnarray}\label{au5}
& T_{\mu \nu }^{ct}=\frac{1}{8\pi G}\frac{2}{\sqrt{-h}}\frac{\delta {{S}_{ct}}}{\delta {{h}^{\mu \nu }}}=T_{\mu \nu }^{1}+T_{\mu \nu }^{F}=\\
& \nonumber \frac{1}{8\pi G} \left[ -\frac{1}{8L{{G}_{3}}}{{h}_{\mu \nu }}-{\alpha}\left[ {{F}_{\mu r}}F_{\nu }^{r}-\frac{{{F}_{\mu r}}{{F}^{r\mu }}}{4}{{h}_{\mu \nu }} \right]\ln \left( \frac{r}{L} \right)\right]
\end{eqnarray}
On the other hand, after renormalization, the expectation value of the stress-energy tensor of the dual $CFT_{2}$,  calculated in \cite{vbala, sdeh}, is defined as follows:
\begin{equation}\label{au4}
{{\left\langle {{T}_{\mu \nu }} \right\rangle }_{CFT}}=T_{\mu \nu }^{ren}
\end{equation}
Using Eqs. (\ref{au6}), (\ref{au5}), and (\ref{au4}),  the $tt$ element of the the stress-energy tensor, ${{\left\langle {{T}_{t t }} \right\rangle }_{CFT}}$, can be expanded up to the second order for each component as in the following relations:
\begin{eqnarray}\label{ww1}
\nonumber \left\langle {{T}_{tt}} \right\rangle_{CFT} \approx  \left\langle T_{tt}^{\left( 1 \right)} \right\rangle_{CFT} +\left\langle T_{tt}^{\left( 2 \right)} \right\rangle_{CFT} = \frac{\kappa }{16\pi GL}\\ 
- \frac{3{{L}^{5}}{{\kappa }^{2}}}{64\pi G{{r}^{2}}}-\frac{{{x}_{3}}L{{\lambda }^{2}}}{8\pi G}\left[ {{y}_{3}}-\ln \left( \frac{r}{L} \right) \right]-\frac{\alpha {{\lambda }^{2}}}{2}\ln \left( \frac{r}{L} \right)
\end{eqnarray}
Here $x_{3}=\left\{ 2,1,1 \right\}$ and $y_{3}=\left\{2, 1, 0\right\}$. We also must choose $\alpha=\frac{L }{8 \pi G} \left\{1,2,2\right\}$ to cancel the divergences of $\left\langle {{T}_{tt}} \right\rangle_{CFT} $ as $r$ approaches  the boundary ($r\to \infty $).
In addition, the energy associated with the subsystem A at the boundary $CFT_{2}$ can be defined as:
\begin{equation}\label{ww2}
E=\int_{-\frac{\gamma }{2}}^{\frac{\gamma }{2}}{dx\left\langle {{T}_{tt}} \right\rangle_{CFT} }
\end{equation}
From Eqs. (\ref{ww1}) and (\ref{ww2}), one could expand the energy as follows:
\begin{equation}\label{s2}
E\approx {{E}^{\left( 0 \right)}}+{{E}^{\left( 1 \right)}}+{{E}^{\left( 2 \right)}}=\frac{\kappa \gamma }{16\pi GL}-\frac{{{x}_{4}}L{{\lambda }^{2}}\gamma }{8\pi G}
\end{equation}
 where $x_{4}=\left\{ 1,1,0 \right\}$ and $E^{(0)}$ is zero for all cases. As shown in Eqs. (\ref{s1}) and (\ref{s2}), it is obvious that the first order of the energy and the  holographic entropy perturbation satisfy a suitable relation analogous to the first law of thermodynamics:
 \begin{equation}
 {{T}_{e}}S_{{{\gamma }_{A}}}^{\left( 1 \right)}={{E}^{\left( 1 \right)}}
 \end{equation}
 where,  ${{T}_{e}}=3/( L^{2}\pi \gamma) $ is the effective temperature (entanglement temperature ) for all the cases.  More precisely, the entanglement temperature is proportional to the inverse of typical size of the
entangling region. It is surprising that we have arrived at an inequality for the second order excitations, namely:
 \begin{equation}
 {{T}_{e}}\left[ S_{{{\gamma }_{A}}}^{\left( 1 \right)}+S_{{{\gamma }_{A}}}^{\left( 2 \right)} \right]<  {{E}^{\left( 1 \right)}}+{{E}^{\left( 2 \right)}}
 \end{equation}
 To tackle this issue and to achieve equality in the first law-like relation when taking into account the second order quantum corrections,
 \begin{equation}
  {{{T}'}_{e}}\left[ S_{{{\gamma }_{A}}}^{\left( 1 \right)}+S_{{{\gamma }_{A}}}^{\left( 2 \right)} \right]= {{E}^{\left( 1 \right)}}+{{E}^{\left( 2 \right)}}
 \end{equation}
   we have to modify the entanglement temperature as follows:
\begin{widetext}
\begin{equation}
{{{T}'}_{e}}={{T}_{e}}\Big[1+\frac{1}{120}\kappa {{\gamma }^{2}}+ \frac{1}{14400}{{\kappa }^{2}}{{\gamma }^{4}}+ \left\{ \begin{array}{cc}
\left[2 L^2 + {b\over 2} \right]\left(-\frac{2{{\lambda }^{2}}}{\kappa }-\frac{{{\lambda }^{2}}{{\gamma }^{2}}}{30}+\frac{4{{\lambda }^{4}}b}{{{\kappa }^{2}}}\right) +\frac{2{{\lambda }^{2}}}{\kappa }L^2 \Big] & BINEF\\
\left[L^2 + a \right]\left(-\frac{2{{\lambda }^{2}}}{\kappa }-\frac{{{\lambda }^{2}}{{\gamma }^{2}}}{30}+\frac{4{{\lambda }^{4}}b}{{{\kappa }^{2}}}\right) +\frac{{{\lambda }^{2}}{{\gamma }^{2}}}{30}L^2 \Big] & ENEF\\
-\frac{2{{\lambda }^{2}}}{\kappa }a-\frac{{{\lambda }^{2}}{{\gamma }^{2}}}{30}a+\frac{4{{\lambda }^{4}}}{{{\kappa }^{2}}}{{a}^{2}} \Big] & LNEF  \\ \end{array} \right.
\end{equation}
\end{widetext}
 where $a=\frac{1}{3}+ln \left( \frac{\gamma}{L} \right)$ and $b=\frac{1}{3}+a$. To conclude, we note that the first law-like relation should also be held at the second order with the modified entanglement temperature. 
 \section{Conclusion}\label{S5}
The entanglement entropy is a powerful tool for describing the entanglement structure of quantum systems. In this paper, we concentrated on the entanglement entropy of the generalized charged BTZ black hole by using the $AdS_{3}/CFT_{2}$ duality.
Following  Ryu and Takayanagi's conjecture \cite{sry, sry2, sry3}, we calculated the bulk 1D-dimensional static minimal
surface up to the second order in terms of electric charge and  mass for the generalized charged BTZ black hole. More precisely, we obtained quantum excitation corrections to the HEE up to the second order by applying gravitational perturbations away from the pure $AdS_3$ spacetime when the spatial region of the boundary CFT is a strip with the length $\gamma$.

Another quantity which can always be defined is the ADM energy of the system. Here, we took account of the energy associated with the subsystem A in the boundary CFT by using the expectation value of the renormalized boundary stress tensor \cite{vbala, kostas, sdeh}. In order to have  the renormalized boundary stress tensor, it is essential to remove the divergences which appear at the boundary by adding local counterterms to the action \cite{vbala, kjensen, kostas}. We obtained the counterterm related to the gauge field which cancels the logarithmic divergence which arises from the gravity action near the boundary \cite{kjensen}.

In addition, combining the variation of energy of the generalized charged BTZ black hole on
the boundary and variation of HEE up to the second order, we showed that the first law-like relation may hold order by order.

\section*{Acknowledgement}
We thank Nima Dehmamy for discussions on all topics presented in this paper, and Nima Lashkari for useful remarks and comments. 



\end{document}